\documentclass[10pt, conference, letterpaper]{IEEEtran}

\usepackage{cite}
\usepackage{amsmath,amssymb,mathtools,float,stfloats,multirow,subfigure,xcolor,amsthm,url}
\usepackage{bm}
\usepackage{upgreek}
\usepackage{algorithm}
\usepackage{algorithmic}
\usepackage{caption}

\usepackage[export]{adjustbox}
\usepackage{graphicx}

\usepackage[T1]{fontenc}
\usepackage[utf8]{inputenc}

 \usepackage{booktabs}

\IEEEoverridecommandlockouts

\begin{document}
\title{SimPyWireless: A PHY-Aware Discrete-Event Simulation Framework for Wi-Fi and 5G NR-U Coexistence in the 5 GHz Band}
\IEEEaftertitletext{\vspace{-0.0\baselineskip}}
\author{\IEEEauthorblockN{Md Rashedur Rahman and Paulo Costa}
	\IEEEauthorblockA{George Mason University, Fairfax, VA, USA\\     
			Email:  [mrahma47, pcosta]@gmu.edu }}
\maketitle

\begin{abstract}

The growing demand from heterogeneous wireless applications with stringent QoS and QoE requirements is increasing pressure on available spectrum resources. Although mmWave and THz bands offer additional capacity, their propagation and hardware limitations make sub-6~GHz unlicensed spectrum, particularly the 5~GHz band, an attractive complementary resource. However, reproducible Wi-Fi/5G NR-U coexistence studies remain limited by the lack of lightweight, open-source, PHY-aware simulation platforms. We present \emph{SimPyWireless}, a discrete-event simulation framework for Wi-Fi and 5G NR-U coexistence in the 5~GHz unlicensed band. The framework integrates shared-channel modeling, propagation, interference, SINR, heterogeneous traffic, mobility, and technology-specific channel access, while providing modular support for externally trained AI/ML models. We validate SimPyWireless against an analytical DTMC model and demonstrate expected coexistence behavior under varying contention and mobility conditions. SimPyWireless provides a reproducible and extensible middle ground between analytical modeling and hardware-based experimentation.


\end{abstract}\vspace{-.05in}




\section{Introduction}\label{sec:Introduction}

\begin{small}\textbf{Background}\end{small}
The growing demand from applications such as massive IoT, industrial automation, and autonomous systems is increasing pressure on available spectrum resources~\cite{alsaedi2023spectrum}. Although higher-frequency bands such as mmWave and THz offer additional spectrum opportunities, their propagation and deployment limitations make sub-6~GHz unlicensed spectrum an important complementary resource~\cite{rahman2024rancad}. 3GPP introduced 5G New Radio-Unlicensed (NR-U) to enable 5G operation in these bands alongside technologies such as Wi-Fi. In the 5~GHz band, NR-U employs Listen-Before-Talk (LBT), introducing contention-based channel access that differs fundamentally from conventional scheduled cellular operation.

\begin{small}\textbf{Motivation}\end{small}
Although Wi-Fi/cellular coexistence in unlicensed spectrum has been extensively studied~\cite{sathya2020standardization,naik2021coexistence,naik2018coexistence}, accessible platforms for reproducible NR-U/Wi-Fi experimentation remain limited. Existing solutions include ns-3/5G-LENA~\cite{giupponi2016simulating,patriciello2020nr}, lightweight SimPy-based models~\cite{zajkac2022resolving}, and experimental implementations using OpenAirInterface and srsRAN~\cite{peixoto2025coexistence,sathish2025xg}. However, existing simulation approaches provide limited PHY-aware coexistence modeling, while experimental platforms require substantial implementation effort and hardware resources. This motivates a lightweight, open-source, and PHY-aware framework for reproducible and extensible Wi-Fi/NR-U coexistence studies.

\begin{small}\textbf{Challenges}\end{small}
Developing an open-source Wi-Fi/5G NR-U experimental platform is challenging because existing cellular frameworks such as OpenAirInterface~\cite{openairinterface5g} and srsRAN~\cite{srsran} are primarily designed for scheduled operation in licensed spectrum. In contrast, NR-U introduces LBT-based channel access with sensing, random backoff, and slot-aligned transmission, requiring substantial MAC/PHY modifications and integration effort. Hardware-based coexistence studies further depend on SDRs, RF components, and controlled test environments, increasing deployment cost and complexity. Simulation offers a more accessible alternative, but accurately modeling Wi-Fi/NR-U coexistence still requires heterogeneous channel access, fine-grained timing, propagation, interference, and mobility within a common framework. These challenges motivate a lightweight and PHY-aware simulation platform for reproducible coexistence studies.

\begin{small}\textbf{Contributions}\end{small}
Motivated by these challenges, we propose \emph{SimPyWireless}, an open-source framework for Wi-Fi/5G NR-U coexistence in the 5~GHz unlicensed band. Our contributions are threefold: (1) a lightweight, PHY-aware simulator that captures channel access, propagation, interference, and mobility, released publicly under the MIT license~\cite{simpywireless}; (2) validation against an analytical DTMC model together with representative coexistence studies involving sensing distance, mobility, and QoS metrics; and (3) a pluggable AI/ML interface for rate adaptation, demonstrated by integrating an externally trained SINR predictor without modifying the core MAC/PHY or channel-access logic.

\begin{small}\textbf{Related Work}\end{small}
Analytical models are widely used to study shared-spectrum access. Bianchi~\cite{bianchi2000performance} introduced a two-dimensional DTMC model for saturated CSMA/CA, later extended to cellular/Wi-Fi coexistence~\cite{gao2016performance,ren2021performance} and adversarial scenarios~\cite{rahman2024rancad,rahman2025channel}. However, such models rely on simplifying assumptions and cannot fully capture time-varying propagation, interference, or mobility. Experimental NR-U implementations using OpenAirInterface~\cite{peixoto2025coexistence} and srsRAN~\cite{sathish2025xg} provide greater realism but require specialized hardware and substantial implementation effort. Simulation offers a practical middle ground: ns-3 supports LTE-U/LAA coexistence~\cite{giupponi2016simulating}, 5G-LENA provides detailed 5G NR simulation~\cite{patriciello2020nr}, and a lightweight SimPy-based Wi-Fi/NR-U simulator was introduced in~\cite{zajkac2022resolving}, though without PHY-layer modeling. In contrast, \emph{SimPyWireless} combines PHY-aware coexistence modeling, mobility, configurable scenarios, and AI/ML extensibility in an open-source framework.

\section{System Model and SimPyWireless Architecture}\vspace{-0.05in} \label{section_sysmodel}


\subsection{Coexistence and Channel-Access Model}
Wi-Fi and 5G NR-U share the same unlicensed channel but use different access procedures: Wi-Fi employs CSMA/CA, while NR-U uses 3GPP Listen-Before-Talk (LBT)~\cite{3GPPrel16}. Both rely on sensing and random backoff; however, NR-U transmissions must align with slot boundaries and may therefore experience a gap after successful LBT. This interval can be handled using a Reservation Signal (RS) or self-deferral, with the latter leaving the channel idle until the next transmission opportunity. Since prolonged RS usage can reduce coexistence fairness~\cite{szott2022using,kosek2021downlink}, SimPyWireless supports both mechanisms.

\subsection{SimPyWireless Architecture}
\begin{figure}[t]
\centering
\includegraphics[width=0.7\linewidth]{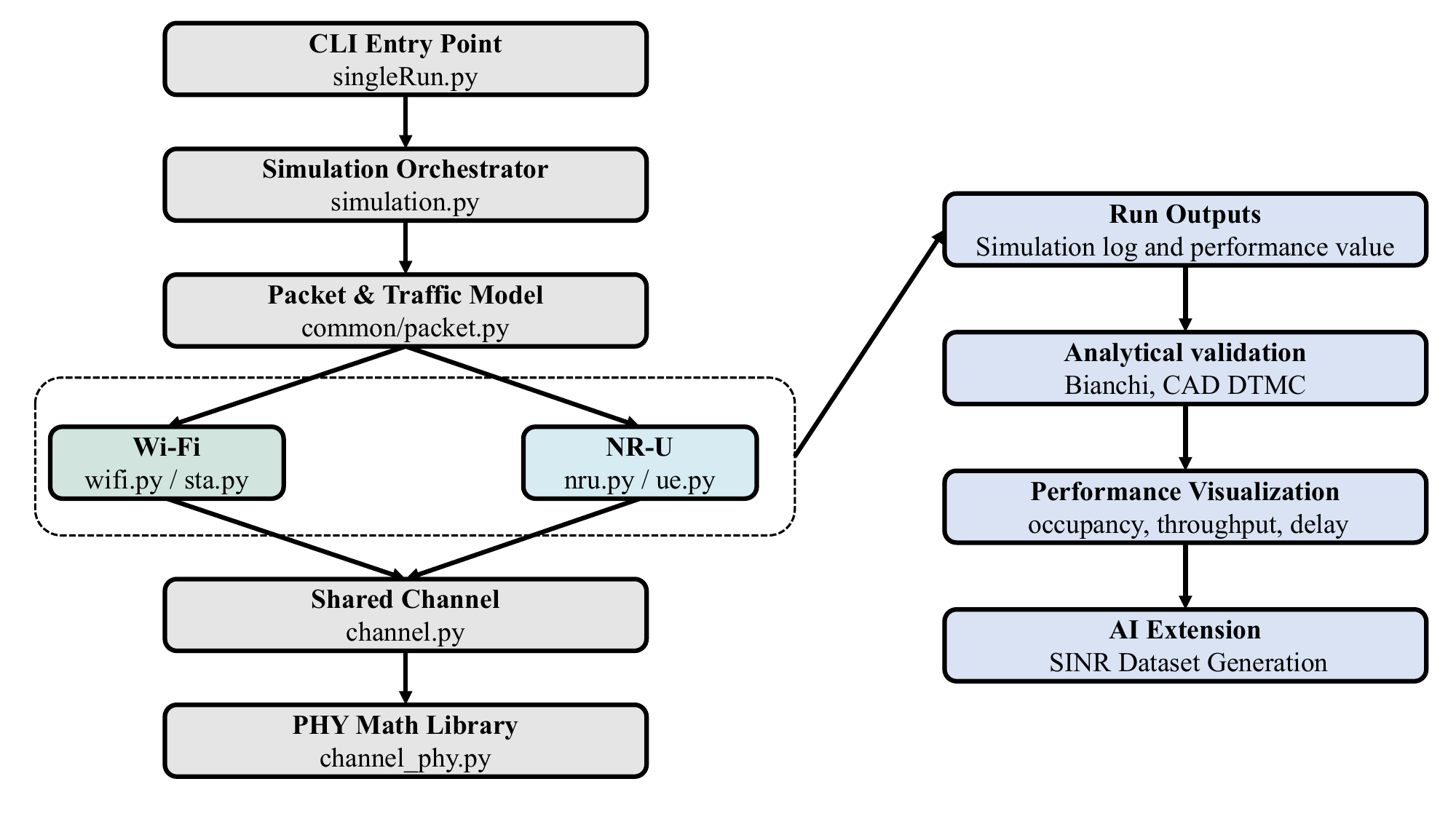}\vspace{-.1in}
     \caption{Simulation Architecture of Simpy Wireless.}
    \label{fig:simulation_architecture}\vspace{-0.1in}
\end{figure}

SimPyWireless is a modular Python/SimPy discrete-event framework comprising configuration, topology, radio-access, PHY, shared-channel, and performance-analysis components, as illustrated in Fig.~\ref{fig:simulation_architecture}. The command-line interface provides a unified entry point for configuring parameters such as Wi-Fi/NR-U node counts, simulation duration, contention windows, transmission power, mobility, traffic, and channel settings, which are passed to the simulation orchestrator for reproducible single runs or parameter sweeps. The orchestrator initializes the simulation environment and shared channel, constructs the topology and traffic sources, executes the discrete-event simulation, and performs post-run analysis. During each run, SimPyWireless maintains packet- and node-level traces and reports key metrics including throughput, channel occupancy, packet delay, packet delivery ratio, and Jain's fairness index~\cite{jain1984quantitative}. Building on earlier SimPy-based coexistence models~\cite{zajkac2022resolving,rahman2024rancad}, the framework further incorporates PHY-aware channel modeling, mobility, heterogeneous traffic, configurable radio-access mechanisms, and extensible AI/ML decision interfaces.


\subsection{Traffic and Radio Access Models}
SimPyWireless uses a technology-agnostic packet and traffic abstraction while preserving technology-specific Wi-Fi and NR-U channel-access behavior. Packets maintain source/destination information, payload size, transmission status, MCS, and per-attempt SINR measurements, enabling direct computation of latency and packet delivery statistics. The framework supports saturated, Poisson, and constant-bit-rate (CBR) traffic, together with IEEE 802.11e/WMM-inspired voice, video, best-effort, and background traffic classes. The Wi-Fi model implements DCF and EDCA, including channel sensing, inter-frame spacing, random backoff, contention-window adaptation, and preamble/energy-based detection. The NR-U model implements 3GPP-style Cat. 3 and Cat. 4 LBT with energy-based CCA, prioritization, random backoff, and configurable MCOT. Since NR-U transmissions must align with eligible slot boundaries, both reservation-signal and self-deferral mechanisms are supported; the evaluations in this work use the gap-based self-deferral configuration.

\subsection{Shared Channel and PHY Model}
SimPyWireless models Wi-Fi and 5G NR-U transmissions over a common wireless channel, where all active transmissions jointly determine channel sensing, interference, and reception outcomes. The default configuration uses a 5.18~GHz, 20~MHz channel, although frequency and bandwidth are configurable. During channel access, received energy is computed from all active transmitters while accounting for spectral overlap, enabling co-channel and partially overlapping interference to be represented. Packet reception is determined from the resulting SINR,

\begin{small}

\begin{equation}
\mathrm{SINR}_{\mathrm{dB}}
=
10\log_{10}
\left(
\frac{P_{\mathrm{rx}}}
{\sum_k I_k + N}
\right),
\end{equation}

\end{small}

\noindent where $P_{\mathrm{rx}}$ is the desired received power, $I_k$ denotes interfering received power, and $N$ is receiver noise. Received power is computed using a configurable log-distance path-loss model with optional log-normal shadowing, and the same propagation model is used for both sensing and reception. SimPyWireless also supports random-waypoint, deterministic linear, and relative mobility models, allowing node movement to dynamically affect path loss, SINR, sensing relationships, and coexistence.\vspace{-0.1in}

\subsection{Experiment Automation and AI/ML Extensibility}
SimPyWireless supports automated parameter sweeps, multiple random seeds, packet-level logging, performance visualization, analytical validation, and regression testing to enable reproducible coexistence experiments. The framework also exposes an optional AI/ML prediction interface through which externally trained models can be integrated into existing communication decisions. In the current implementation, a predictor can replace the most recently observed SINR used for rate adaptation, while SINR measurements from both successful and failed transmission attempts are retained for offline training and evaluation. The interface remains inactive by default and allows different prediction models to be evaluated without modifying the core MAC, PHY, or shared-channel.

\section{Validation and Simulation Capabilities}

In this section, we first validate SimPyWireless against an analytical model of Wi-Fi and 5G NR-U coexistence in the 5~GHz unlicensed band. We then present representative case studies to demonstrate the simulator's ability to capture heterogeneous coexistence behavior, mobility-driven channel dynamics, and AI/ML-assisted communication mechanisms.\vspace{-.05in}

\subsection{Mathematical Validation}
\begin{figure}[!t]
    \centering
    {\includegraphics[width=0.2\textwidth]{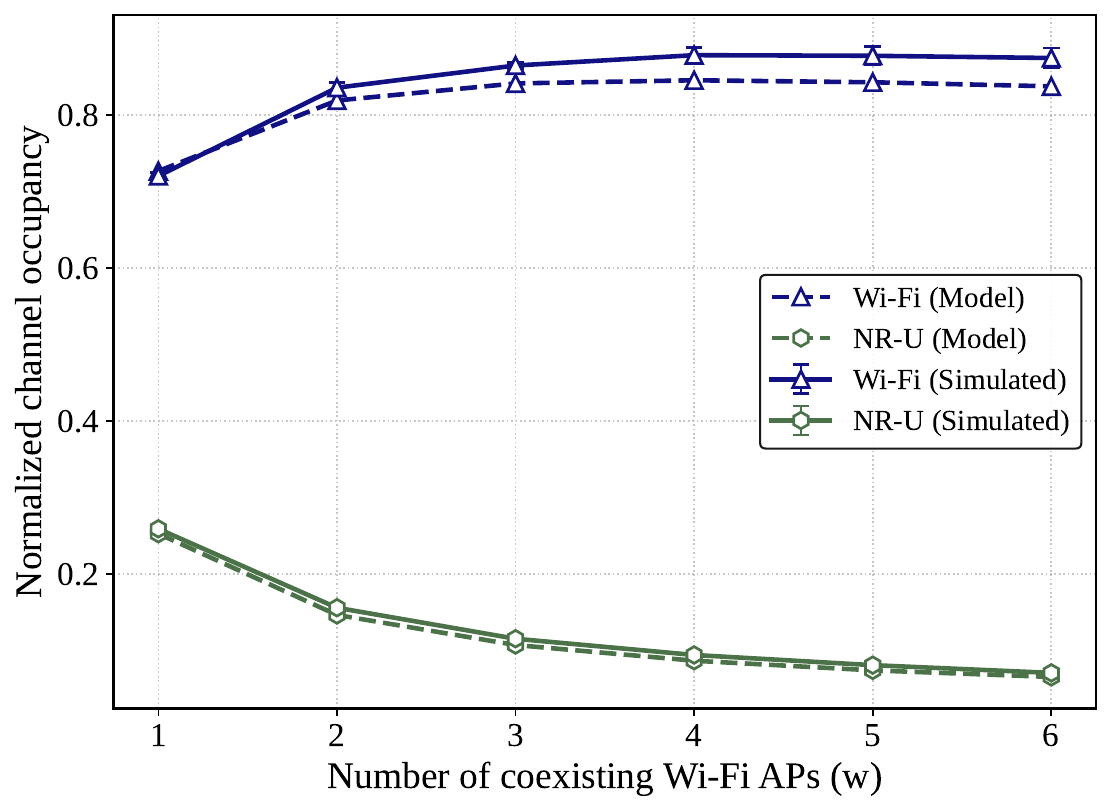}}
    \caption{\footnotesize Validation of SimPyWireless against the DTMC model in terms of normalized Wi-Fi and NR-U channel occupancy.}\vspace{-.1in}
    \label{fig:simulation_vs_model}
\end{figure}
To validate the fidelity of SimPyWireless, we compare its normalized channel-occupancy results against the DTMC-based analytical formulation adopted from~\cite{rahman2025channel,gao2016performance,ren2021performance}. As shown in Fig.~\ref{fig:simulation_vs_model}, the simulation closely follows the analytical predictions across all evaluated Wi-Fi densities. As the number of coexisting Wi-Fi APs increases from one to six, aggregate Wi-Fi channel occupancy increases while the corresponding NR-U occupancy decreases. The agreement is particularly close under lower contention levels, with modest deviations appearing as the number of Wi-Fi contenders increases. Overall, SimPyWireless preserves both the magnitude and scaling behavior predicted by the analytical model, providing confidence in its implementation.

\begin{figure}[t]
    \centering
    \subfigure[\footnotesize Channel Occupancy]{\includegraphics[width=0.2\textwidth]{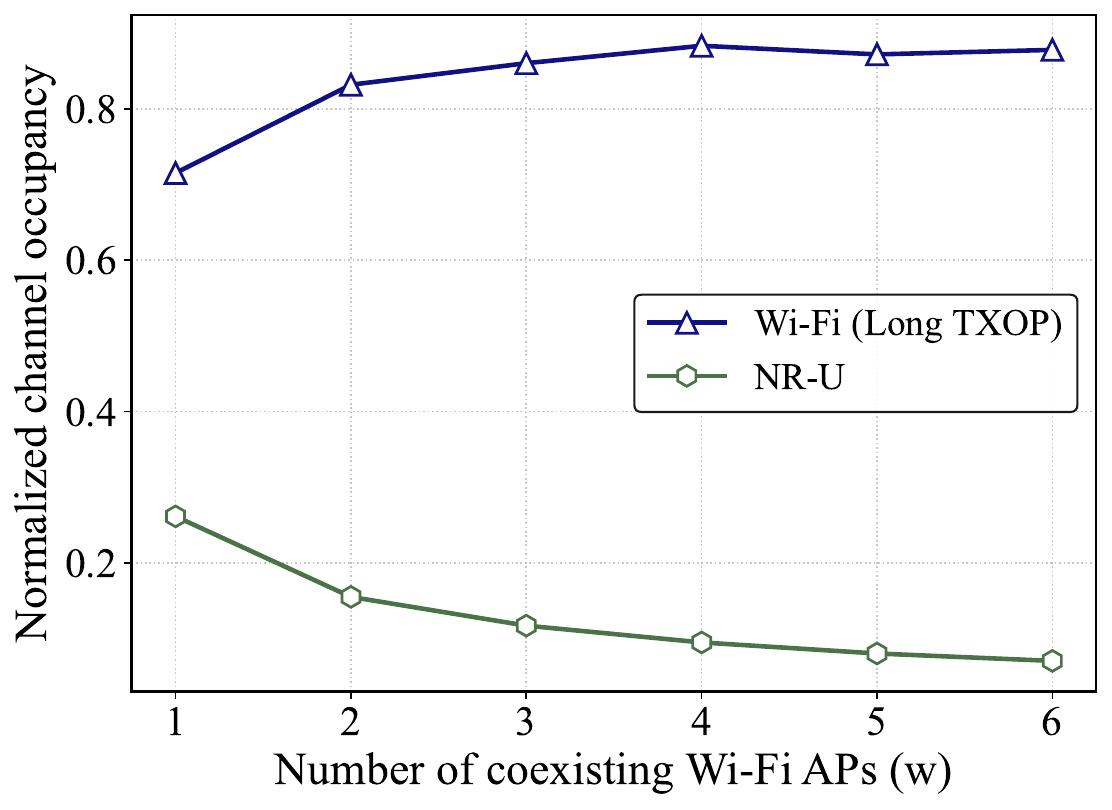}}
    \subfigure[\footnotesize Throughput]{\includegraphics[width=0.2\textwidth]{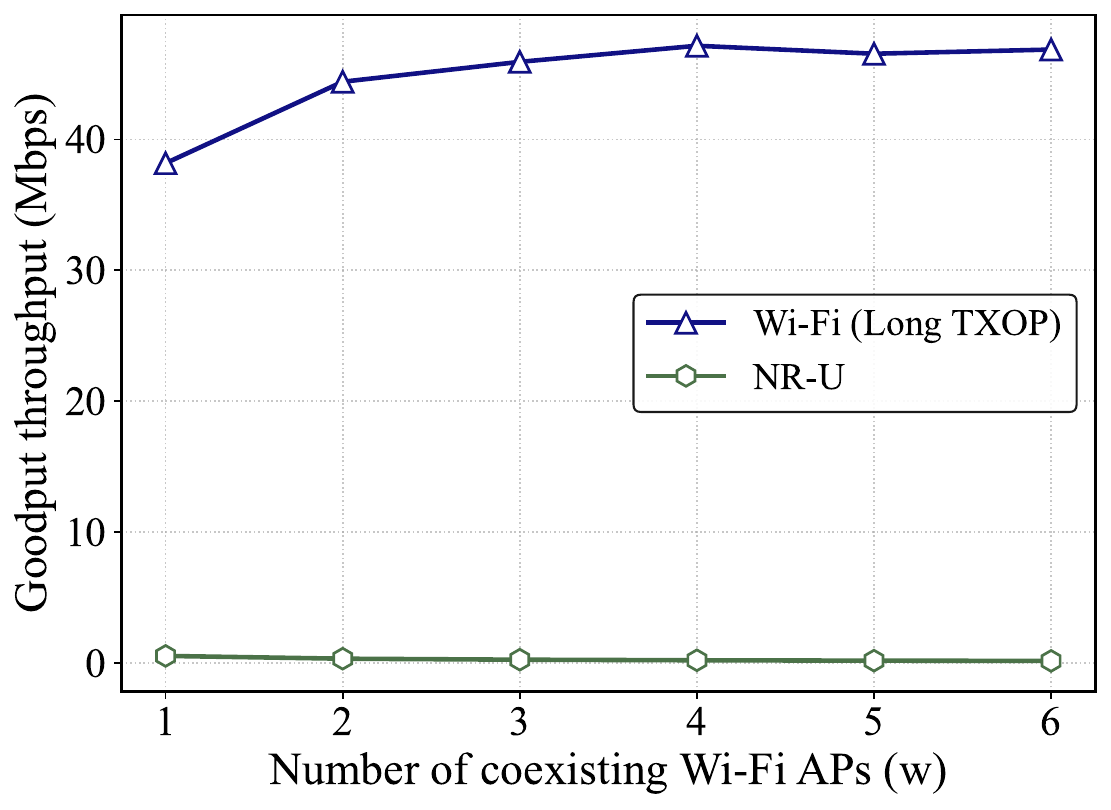}}\\
    \subfigure[\footnotesize Delay]{\includegraphics[width=0.2\textwidth]{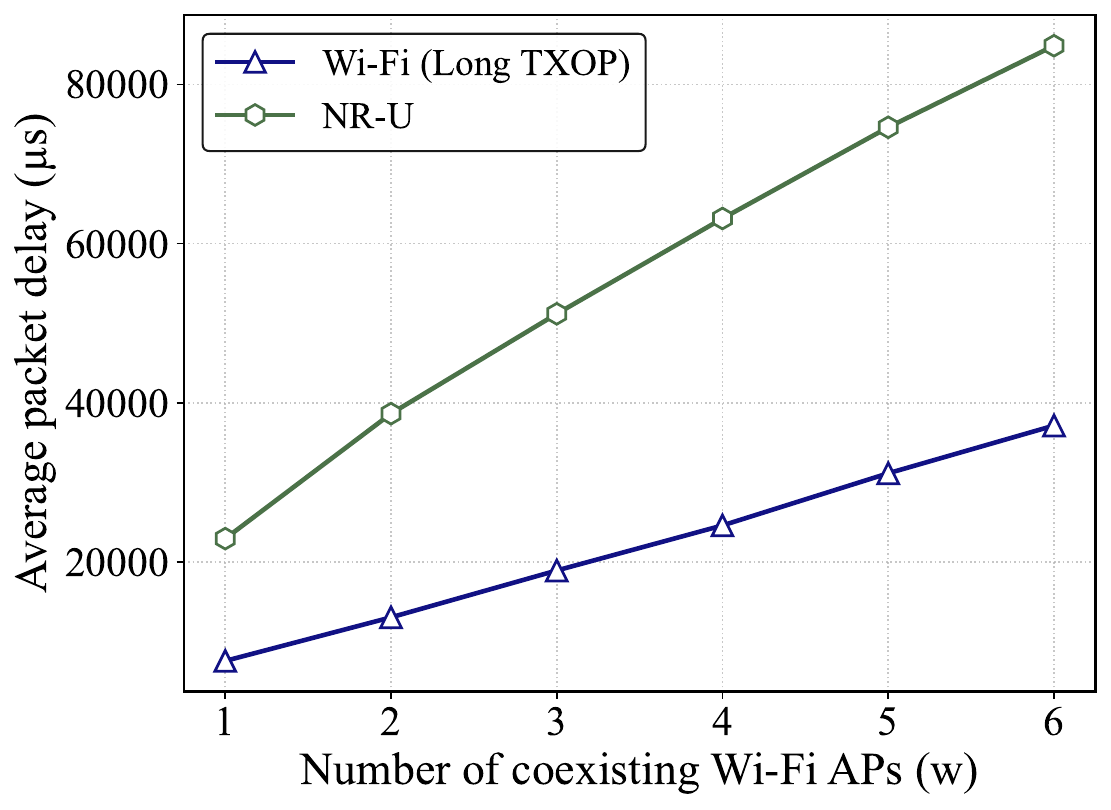}}
    \subfigure[\footnotesize Fairness]{\includegraphics[width=0.2\textwidth]{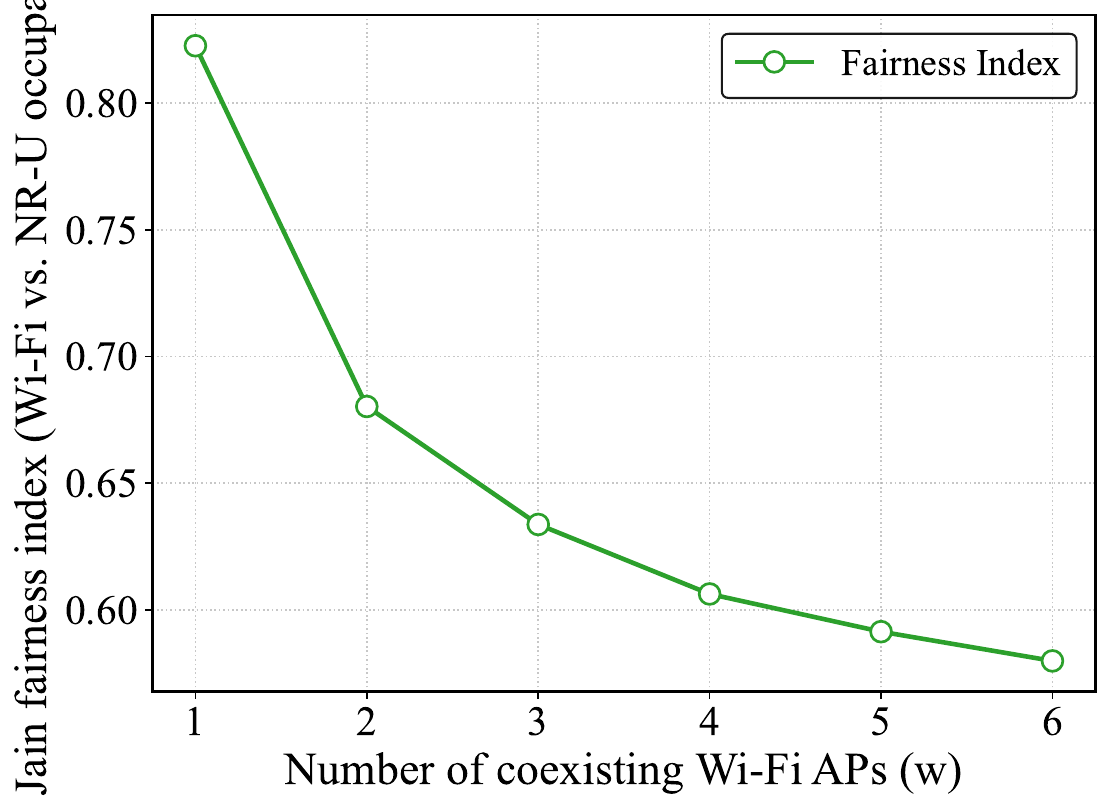}}

    \caption{\footnotesize Performance of Wi-Fi and 5G NR-U coexistence with an increasing number of coexisting Wi-Fi APs.}\vspace{-0.1in}
    \label{fig:simu_performance_analysis_wifi_nru}
\end{figure}

\begin{figure}[t]
    \centering
    {\includegraphics[width=0.2\textwidth]{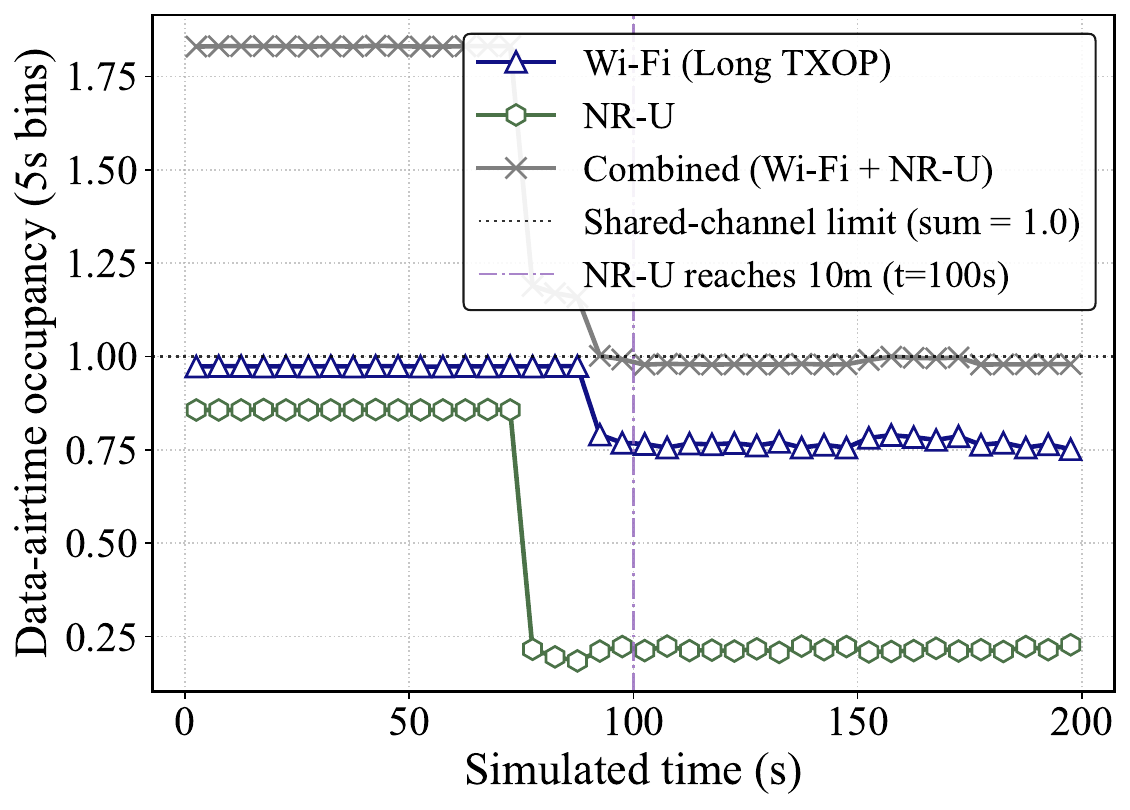}}
    \caption{\footnotesize Evolution of Wi-Fi and 5G NR-U channel occupancy under mobility, illustrating the transition from independent to shared-channel operation.}\vspace{-.1in}
    \label{fig:mobility_analysis}
\end{figure}

\subsubsection{QoS Performance}
To evaluate coexistence performance, we conduct $200$-s simulations in which a single NR-U gNB coexists with one to six Wi-Fi APs, with all nodes located within mutual sensing range. As shown in Fig.~\ref{fig:simu_performance_analysis_wifi_nru}(a), increasing the number of Wi-Fi APs raises aggregate Wi-Fi channel occupancy from approximately $0.72$ to $0.87$, while NR-U occupancy decreases from about $0.26$ to $0.07$. A similar trend appears in throughput [Fig.~\ref{fig:simu_performance_analysis_wifi_nru}(b)], where NR-U throughput declines while aggregate Wi-Fi throughput increases and saturates near $47$~Mbps. Increased contention also raises packet delay for both technologies, with a larger degradation for NR-U [Fig.~\ref{fig:simu_performance_analysis_wifi_nru}(c)]. Consequently, Jain's fairness index decreases from approximately $0.82$ to $0.58$ [Fig.~\ref{fig:simu_performance_analysis_wifi_nru}(d)]. These results demonstrate SimPyWireless's ability to jointly characterize performance metrics under heterogeneous coexistence.

\subsubsection{Mobility Analysis}
To demonstrate mobility-aware coexistence modeling, we consider a scenario in which a Wi-Fi AP and an NR-U node move toward each other and eventually enter mutual sensing range. As shown in Fig.~\ref{fig:mobility_analysis}, both technologies initially maintain relatively stable channel occupancy while sufficiently separated. As the nodes approach one another, the transition to shared-channel contention becomes visible between approximately $75$ and $90$~s. Once mutual sensing is established, NR-U experiences a larger reduction in airtime than Wi-Fi. This transition demonstrates the simulator's ability to capture mobility-induced changes in sensing relationships, interference, and channel-access behavior within an experiment.\vspace{-0.05in}

\subsection{AI/ML Extensibility}
Rather than proposing a new AI/ML algorithm, this study uses a trained SINR predictor as a representative example to demonstrate how externally developed learning models can be integrated into SimPyWireless without modifying its core MAC, PHY, or channel-access logic. The predictor is inserted into the existing rate-adaptation pipeline, allowing the simulator to evaluate both prediction accuracy and its subsequent impact on communication performance within the same coexistence environment.

\subsubsection{SINR Prediction}
\begin{figure}[t]
    \centering
    \subfigure[\footnotesize MAE]{\includegraphics[width=0.2\textwidth]{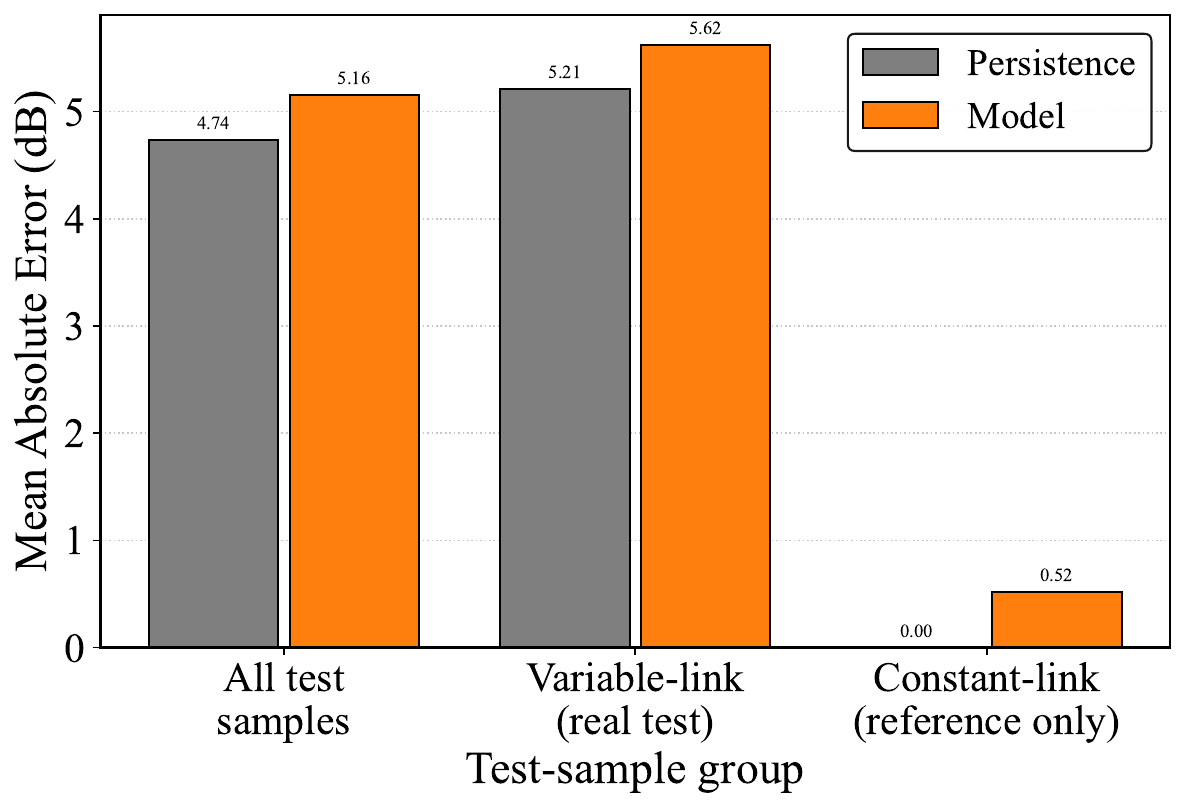}}
    \subfigure[\footnotesize RMSE]{\includegraphics[width=0.2\textwidth]{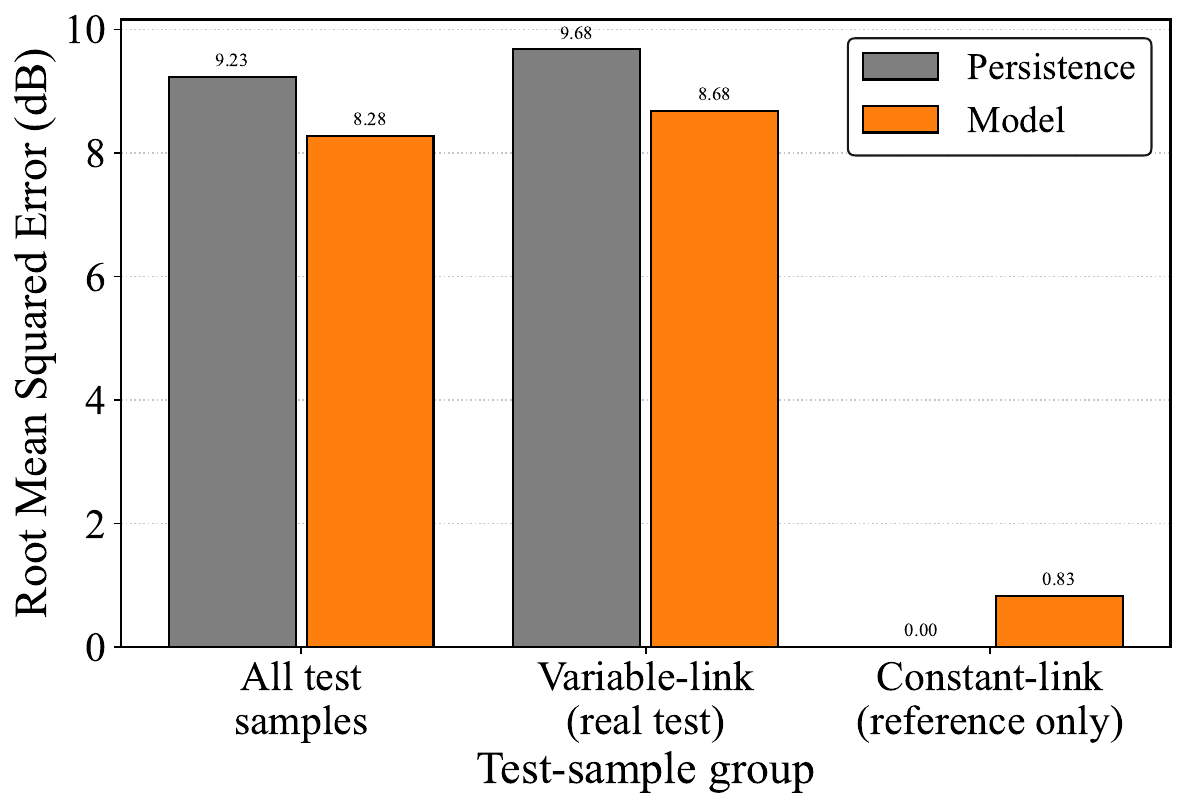}}\\
    
    \caption{\footnotesize SINR prediction error of a model plugged into the simulator's predictor hook, vs.\ a persistence baseline}\vspace{-0.2in}
    \label{fig:ml_analysis_simpy_sinr}
\end{figure}

To demonstrate support for data-driven channel-quality estimation, we evaluate a trained gradient-boosting SINR predictor against a persistence baseline that predicts the next SINR from the most recent observation. As shown in Fig.~\ref{fig:ml_analysis_simpy_sinr}, the learned model achieves lower RMSE for both the complete test set and the variable-link subset, reducing error from $9.23$ to $8.28$~dB and from $9.68$ to $8.68$~dB, respectively. In contrast, the persistence baseline achieves slightly lower MAE. These results indicate that the learned predictor better limits larger prediction deviations, while also illustrating that SimPyWireless can integrate externally trained models within its existing decision pipeline.\vspace{-0.05in}

\subsubsection{Impact on Packet Delivery Ratio}
\begin{figure}[t]
    \centering
    {\includegraphics[width=0.2\textwidth]{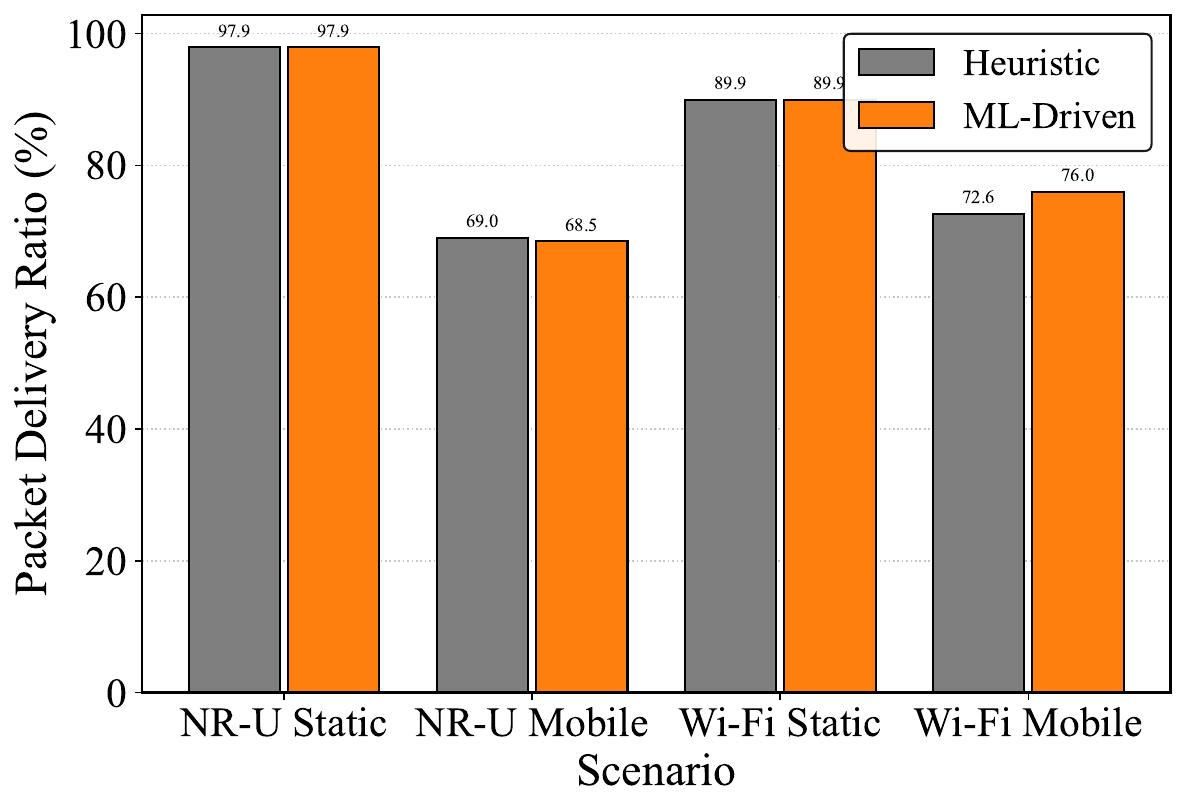}}
    \caption{\footnotesize Packet delivery ratio with the trained SINR predictor plugged into the simulator's rate-adaptation hook, vs.\ the default heuristic}
    \label{fig:ml_pdr_comparison}
\end{figure}

To evaluate the end-to-end impact of the learned SINR predictor, we compare packet delivery ratio (PDR) against the simulator's default rate-adaptation mechanisms, namely ARF for Wi-Fi and CQI-based adaptation for NR-U. As shown in Fig.~\ref{fig:ml_pdr_comparison}, both approaches perform similarly under static conditions. Under mobility, NR-U remains nearly unchanged, with PDR values of $69.0\%$ and $68.5\%$ for the heuristic and ML-driven approaches, respectively, while Wi-Fi improves from $72.6\%$ to $76.0\%$. These results show that the impact of learned channel-quality prediction is scenario-dependent while demonstrating the framework's ability to evaluate AI/ML-assisted communication strategies end-to-end.

\textbf{Summary:} The analytical validation and representative case studies demonstrate that SimPyWireless can reproduce expected Wi-Fi/NR-U coexistence behavior while supporting QoS analysis, mobility-driven transitions, and data-driven extensions. Together, these results highlight the framework's suitability for reproducible coexistence studies without requiring a dedicated hardware testbed, while also demonstrating its extensibility toward AI/ML-assisted wireless research.

\section{Conclusion}
This work presents \emph{SimPyWireless} as an open-source, PHY-aware simulation framework for studying Wi-Fi and 5G NR-U coexistence in the 5~GHz unlicensed band. By incorporating propagation, interference, and SINR, the framework captures heterogeneous wireless interactions beyond purely analytical models while avoiding the complexity of hardware-based experimentation. Its configurable architecture supports diverse mobility patterns, traffic models, and application-oriented scenarios within a common environment. SimPyWireless is validated against an existing DTMC-based analytical model, showing close agreement in coexistence behavior and supporting its reliability for Wi-Fi/NR-U studies. Overall, the framework provides a reproducible middle ground between mathematical analysis and experimental testbeds. Future work will extend SimPyWireless toward additional spectrum bands, richer PHY-layer features, multi-channel operation, and application-specific coexistence scenarios.

\bibliographystyle{IEEEtran}
\bibliography{INFOCOM_2020_ref}



\end{document}